\documentclass[twoside,twocolumn,pra]{revtex4}

\usepackage[utf8]{inputenc} 
\newcommand{\titel}{Mathematical model of interest matchmaking in electronic social networks}
\usepackage{amsmath}  
\usepackage{amssymb}
\usepackage{amsthm}
\usepackage{mathrsfs}

\usepackage{color}
\usepackage{eurosym}
\usepackage{makeidx}
\usepackage{verbatim}

\usepackage{ifpdf}
\ifpdf
\usepackage[pdftex]{graphicx}
\usepackage[pdftitle={\titel},pdfauthor={Andreas de Vries},%
            pdfpagemode=UseNone,%
            colorlinks=false,hyperindex=true,plainpages=false,%
           ]{hyperref}
           \DeclareGraphicsExtensions{.pdf,.jpg,.png}
\else
\usepackage{graphicx}
\usepackage[pdftitle={\titel},pdfauthor={Andreas de Vries},%
            colorlinks=false,hyperindex=true,plainpages=false,%
            dvips=true,ps2pdf=true%
           ]{hyperref}
           \DeclareGraphicsExtensions{.eps,.EPSF}
\fi

\newcommand{\URL}  [1]{\textit{#1}}

\newcommand{\pinkverb}[1] {\color[rgb]{.8,.2,.4}}
\newcommand{\schwarzverb}[1] {\color[rgb]{0,0,0}}

\newcounter{uebnr}

\newcounter{merknr}

\newcounter{ratnr}

\newtheorem{satz}{Theorem}

\newtheorem{defi}[satz]{Definition}

\newtheorem{rem}[satz]{Remark}
\newtheorem{beisp}[satz]{Example}

\newtheorem{beispe}[satz]{Examples}
\newtheorem{exercise}[satz]{Exercise}

\newtheorem{algorithm}[satz]{Algorithm}
\newenvironment{definition} {\begin{defi}\begin{em}}{\end{em}\hfill{$\square$}\end{defi}}
\newenvironment{beispiel} {\begin{beisp}\begin{em}}{\end{em}\hfill{$\square$}\end{beisp}}

\newenvironment{pgm*} {\begin{tt}
                        \begin{footnotesize}
                          \begin{tabular}{l}
                      }
                      {    \\
                        \end{tabular}
                        \end{footnotesize}
                       \end{tt}}

\newenvironment{boxedpgm*} {\begin{tt}
                        \begin{footnotesize}
                          \begin{tabular}{|l|}
                          \hline}
                     {    \\ \hline
                        \end{tabular}
                        \end{footnotesize}
                       \end{tt}
                     }
\newenvironment{pseudo*} {\\ \\
                         \begin{em}
                            \begin{tabular}{l}
                        }
                        {     \\ \phantom{+}
                            \end{tabular}
                         \end{em} \\}
\newenvironment{boxedpseudo*} {
                         \begin{em}
                            \begin{tabular}{|l|}
                            \hline
                        }
                        {     \\ \hline
                            \end{tabular}
                         \end{em}}

\makeindex

\begin{document}

\author{Andreas de Vries}%
\thanks{
	South Westphalia University of Applied Sciences,
	Haldener Stra{\ss}e 182,
	D-58095 Hagen, Germany,
	e-Mail: 
	\href{mailto:de-vries@fh-swf.de}{de-vries@fh-swf.de}%
}
\title{\titel}
\date{\today}


\begin{abstract}
	\noindent
	The problem of matchmaking in electronic social networks
	is formulated as an optimization
	problem. In particular, a function measuring the matching degree
	of fields of interest of a search profile with those of an advertising profile
	is proposed.
	
	\medskip
	\noindent
	\textbf{\href{http://www.acm.org/class/1998/}{ACM Classification:}}
	I.2.4: Semantic networks, 
	H.3.5: Web-based services,  
	J.4: Sociology, 
	G.2.3: Applications of discrete mathematics,  
	C.2.4: Distributed applications
	
	
\end{abstract}

\maketitle


\section{Introduction}
Social activities in electronic networks 
play an increasingly important role in our every-day lives.
We are exchanging important information via electronic mails, wikis, web-based forums,
or blogs, and meet new friends or business contacts in Internet communities
and social network services.
Parallel to this growing socialization of the World Wide Web, 
the requirements on the electronic services
become more ambitious. Huge data quantities have to be processed,
user-friendly interfaces are to be designed,
and more and more sophisticated computations must be implemented
to offer complex solutions.

This paper studies a special aspect of social network services, the
matchmaking problem.
In essence it asks, given a search profile, for advertising profiles matching
it best. This problem is in principle
well-known in Grid computing, where computational
tasks 
are seeking for appropriate resources such as CPU time and
memory space on different computers.
In electronic social networks, however, the problem is more general because
not only specified attribute ranges 
are to be compared
but more or less vaguely describable interests.

The aim of this paper is to formulate a mathematical model for the problem
of matchmaking of attribute ranges and fields of interests in electronic social networks.
It tackles the following fundamental questions.
How can an appropriate system and its data structure be designed?
How is the mathematical formulation of a matching problem as an optimization problem?
In particular, what is its search space, what is its objective function?
Whereas the idea to use a fuzzy function to calculate the matching degree of
two numerical ranges may suggest itself, 
how could a function
calculating a matching degree of two fields of interest look like?
One of the central results of this paper is the proposal of a precise
definition of such a function computing this matching degree 
and the presentation of a concrete example.

The paper is organized as follows.
After a definition of electronic social networks is given in the next section,
a mathematical model of the matchmaking problem as an optimization problem is proposed,
especially the data structure of search and advertising profiles,
the search space, and the matching degree as the objective function.
A short discussion concludes the paper.

\section{Electronic social networks}
A \emph{social network}\index{social network} 
consists of a finite set of actors and the
direct relations defined on them. 
An actor here may be an individual, a group, or an organization,
and 
the direct relation between two actors indicates that they
directly interact with each other, have immediate contact,
or are connected through social 
familiarities, such as casual acquaintance or familial bonds
\cite{Barnes-1954,Wasserman-Faust-1994}.
Thus a social network is naturally represented by a graph in which
each node represents an actor and each edge a direct relation.
Empirically, 
the mean number of
direct relationships of an individual in a biological
social network depends on the size of the neocortex of its individuals;
the maximum size of such relationships in human social networks tends to 
be around 150 people (``Dunbar's number''\index{Dunbar's number})
and the average size around 124 people \cite{Hill-Dunbar-2002}.

Since the popularization of the World Wide Web in the middle of the 1990's,
there emerged several Internet social networks, maintained by
social network services such as
``circle of friends'' like
friendster (\href{http://www.friendster.com}{\URL{www.friendster.com}}), 
MySpace (\href{http://www.myspace.com}{\URL{www.myspace.com}}),
or
orkut (\href{http://www.orkut.com}{\URL{www.orkut.com}}),
as platforms for business professionals like
XING (\href{http://www.xing.com}{\URL{www.xing.com}}),
or virtual worlds like Second Life (\href{http://www.secondlife.com}{\URL{secondlife.com}}).
Internet social networks are instances of electronic social networks.

In this paper, an \emph{electronic social network}\index{electronic social network}
is defined as 
a network of at least three human individuals or organizations which use essentially, 
albeit not exclusively,
electronic devices and media 
to get in contact and acquaintance, to meet new partners,
to communicate, and to exchange information.
Examples of electronic social networks are Internet social networks,
as well as videoconference sessions and
conference calls, especially if they serve to meet new people
as in party lines, or as long as they admit spontaneous
communication between each member of the network.


\section{The matchmaking problem}
In computer science, the term \emph{matching}\index{matching} in general refers to the
process of evaluating the degree of similarity or of agreement of two objects.
Each object is characterized by a set of properties or attributes,
which in many systems are given by name-value pairs 
\cite{de-Vries-2004-xml}.
Matching plays a vital role in many areas of computer science and
communication systems. For instance, 
it is studied for resource discovery and resource allocation in grid computing
where matchmaking services are needed to intermediate between resource requesters
and resource providers \cite{Bai-et-al-2004}.
Other examples are given by the problem of matching demands and supply of business
or personal profiles 
in online auctions, e-commerce, 
recruitment agencies, or dating services.

\subsection{Profiles}
In most matching problems, the objects under consideration take asymmetric roles,
viz., some try to \emph{search} for information or {request} for a service,
others try to \emph{advertise} information or {provide} a service.
A single object may naturally do both activities at a time, 
in electronic social networks this even is the usual case.
In the sequel we will therefore more accurately consider the matching of
a \emph{search profile}, containing the requested information,
and an \emph{advertising profile} presenting the provided information.

Given a specific search profile,
the \emph{matchmaking problem}\index{matchmaking problem} then is to
find those advertising profiles which match it best,
in a sense to be specified in the sequel.
Generalizations of this problem ask for best global matchings, 
given a whole set of search profiles 
and a set of advertising profiles. 
For instance, the 
\emph{global pairwise matchmaking problem}\index{global pairwise matchmaking problem}
seeks pairs of search/advertising-profiles
such that the entity of the pairs matches the best under the constraint
that any profile is member of at most one pair,
the 
\emph{global multiple matchmaking problem}\index{global multiple matchmaking problem}
searches for possibly multiple combinations of search and advertising profiles 
which as a whole match the best.
The pairwise version of the problem typically occurs for dating services or classical
marriage matchmaking tasks, whereas the multiple version appears
in grid computing or in brokering interest groups.

In this paper we will focus on the local version of the matchmaking problem, i.e., 
finding an optimum advertising profile to a specified search profile.
Thus the matchmaking problem is an optimization problem, and 
to formulate it precisely we have to specify the search space and the
objective function.
The search space will turn out to be the set of pairs of 
the fixed search profile and the advertising
profiles, and the objective function will be a function measuring the ``matching degree.''
We will work out these notions in the next sections.

\subsubsection{Definitions}
\begin{figure}[htp]
\centering
\begin{footnotesize}
 \begin{tabular}{|c|}
   \hline 
	\textbf{Owner}         \\[.5ex] 
	\hline 
	\multicolumn{1}{|l|}{id: String}\\
	\multicolumn{1}{|l|}{height: {integer}} \\
	\multicolumn{1}{|l|}{eye color: String} \\
	\multicolumn{1}{|l|}{\ \ldots} \\
 	\hline 
 \end{tabular}%
    \hspace*{-.5ex}%
    \begin{tabular}{c}
      \\[-5.0ex]
      1 \hspace*{1.8ex} $*$\\[-1.5ex]
      ---\hspace*{-.25ex}---\hspace*{-.25ex}---
    \end{tabular}%
    \hspace*{-.5ex}%
 \begin{tabular}{|c|} 
   \hline 
	\textbf{Profile}         \\[.5ex] 
	\hline
	\\
	\multicolumn{1}{|l|}{} \\
	\hline 
 \end{tabular}%
    \hspace*{-.5ex}%
    \begin{tabular}{c}
      \\[-5.0ex]
      1 \hspace*{1.8ex} $*$\\[-1.3ex]
      \begin{tabular}{@{}c@{}}\\[-2.6ex]%
     	  \includegraphics[width=2.5ex]{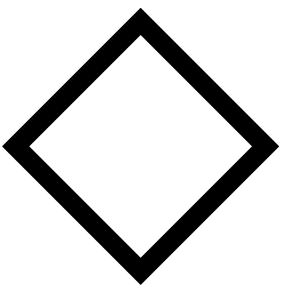}%
      \end{tabular}%
      ---\hspace*{-.25ex}---
      \\[2.0ex]
      1 \hspace*{1.8ex} $*$\\[-1.3ex]
      \begin{tabular}{@{}c@{}}\\[-2.6ex]%
     	  \includegraphics[width=2.5ex]{raute}%
      \end{tabular}%
      ---\hspace*{-.25ex}---
    \end{tabular}%
    \hspace*{-.5ex}%
 \begin{tabular}{|c|} \hline 
	\textbf{Interest}         \\[.5ex] 
	\hline 
 	\multicolumn{1}{|l|}{field: String} \\
 	\multicolumn{1}{|l|}{level: $[-1, 1]$} \\
 	\hline
	\multicolumn{1}{c}{} \\
	\hline
	\textbf{Stencil}  \\[.5ex]
	\hline
 	\multicolumn{1}{|l|}{attribute: String} \\
 	\multicolumn{1}{|l|}{range: $\mathscr{T}$} \\
	\hline
 \end{tabular}
\end{footnotesize}
	\caption{\label{fig-uml-matchmaking}\footnotesize
		UML diagram of the data structure of a profile 
		and its relationship to the owner's attributes,
		the attribute stencils and the fields of interest.
		An attribute stencil consists of an owner's attribute name
		and its (searched or advertised) range.
	}
\end{figure}

A \emph{profile}\index{profile} consists of its owner,
being an actor of the electronic social network,
a list of 
\emph{attributes}\index{attribute} of a given set $A$
together with their values,
a list of \emph{attribute stencils}
where each stencil represents a pair of an attribute name and its
value range,
and a list
of \emph{fields of interest}\index{field of interest}\index{interest}
specifying their respective levels of interest.
Attributes are properties of the profile owner
such as age, height, weight, eye color, or hair color.
and we therefore subsume them under the class ``Owner''
(Fig. \ref{fig-uml-matchmaking}).
In principle, there are two different types of attributes,
subsumed in the two disjoint sets $N$ and $D$ such that the
set $A$ of attributes separates as
\begin{equation}
	A = N \cup D.
\end{equation}
The set $N$ consists of the \emph{numerical attributes}\index{numerical attribute} 
of the owner which take integer or real numbers as values,
the set $D$ consists of discrete non-numerical values.
(The difference between numerical and non-numerical is not sharp,
for instance a string could well be considered as numerical via a
symbol code, as well as non-numerical since it seldom makes sense to
multiply or divide strings; in most cases, strings are better considered
as non-numerical.)

Correspondingly, the stencil of an attribute is determined by the
attribute's name and its range, being of a certain set called
\emph{Type} $\mathscr{T}$,
\begin{equation}
	\mathscr{T} = \mathscr{N} \cup \mathscr{D}
	.
\end{equation}
where $\mathscr{N}$ denotes the set of ranges for the numerical attributes,
\begin{equation}
	\mathscr{N} \subset \{[a,b]: -\infty \leqq a, b \leqq \infty\},
\end{equation}
i.e., $\mathscr{N}$ is a set of closed intervals $[a,b] \subset \mathbb{R}$, 
and $\mathscr{D}$ denotes the set of ranges for the discrete non-numerical attributes,
\begin{equation}
	\mathscr{D} \subset \{E: \mbox{$E$ is a finite set or enum}\}
	,
\end{equation}
i.e., $\mathscr{D}$ is a finite set or enum, specified by the respective
owner attributes determined by the system model.
We allow the empty set $\emptyset$ as null element in $\mathscr{N}$ and
$\mathscr{D}$.
If a given range $R$ $\in$ $\mathscr{T}$ contains only one element, say $R=\{x\}$,
then the stencil is often shortly written ``$p=x$'' instead of ``$p \in R$.''
If, on the other hand, $R=[x,\infty]$ then we may write ``$p>x$'' instead
of $p\in R$.
For instance, 
``height = 180'' means ``height $\in$ $[180, 180]$,''
or 
``height > 180'' means ``height $\in$ $[180, \infty]$.''

On the other hand, a field of interest is a name-value pair specifying the field
itself as well as its level ranging on a scale from $-1$ to 1,
coded by the interpolation of the following table,
\begin{equation}
	\mbox{%
		\begin{tabular}{|r|l|}
			\hline
			\textbf{Level} & \textbf{Meaning}
			\\ \hline
			$-1$ \ & aversion \\
			0    \ & indifference \\
			1    \ & enthusiasm
			\\ \hline
		\end{tabular}
	}
	\label{eq-interest-coding}
\end{equation}
The set of fields of interests is denoted by $I$ and is a subset of words of a
specified alphabet $\Sigma$,
\begin{equation}
	I \subseteq \Sigma^*
\end{equation}
Usually, $\Sigma$ is the set of ASCII or Unicode symbols.
The set $I$ determines the set of all fields of interests 
available to the system. Depending on the system design, it may be
a fixed set of words, or an arbitrary word over the alphabet $\Sigma$.

\subsection{The search space}
Given a set $\mathfrak{S}$ of search profiles $\mathfrak{s}$
and a set $\mathfrak{A}$ of advertising profiles $\mathfrak{a}$
as input,
the search space $S$ of a \emph{global} matchmaking problem is given by
all pairs of search and advertising profiles, i.e.,
$ 
	S^{\mathrm{g}} =
	\mathfrak{S} \times \mathfrak{A}.
$ 
In this paper, however, we are considering the local matchmaking problem,
given a single search profile $\mathfrak{s}$, i.e., $\mathfrak{S} = \{\mathfrak{s}\}$,
and the search space
\begin{equation}
	S =
	\{\mathfrak{s}\} \times \mathfrak{A}(\mathfrak{s}).
	\label{eq-matchmaking-search-space}
\end{equation}
where 
$
	\mathfrak{A}(\mathfrak{s})
	= \{\mathfrak{a} \in \mathfrak{A}: 
	\mbox{owner($\mathfrak{a}$) $\ne$ owner($\mathfrak{s}$)}\}
	.
$
A search profile $\mathfrak{s}$ itself is a set
of the given attribute stencils $\mathfrak{n}_{\mathrm{s}}$, $\mathfrak{d}_{\mathrm{s}}$, 
and of fields of interest $\mathfrak{i}_{\mathrm{s}}$,
\begin{equation}
	\mathfrak{s}
	= \mathfrak{n}_{\mathrm{s}} \cup \mathfrak{d}_{\mathrm{s}} \cup \mathfrak{i}_{\mathrm{s}}, 
	\label{eq-search-profile}
\end{equation}
where 
$$
	\mathfrak{n}_{\mathrm{s}} 
	= \{(p, R_{\mathrm{s}}(p)): p \in N_{\mathrm{s}}\}
$$
is the set of attribute-range pairs, with the given mapping
$R_{\mathrm{s}}: N_{\mathrm{s}} \to \mathscr{N}$
from the set $N_{\mathrm{s}}$ of the searched numerical attributes
to their associated desired ranges
($R_{\mathrm{s}}$ associates to each numerical attribute $p$ in 
$N_{\mathrm{s}}$ an interval $R_{\mathrm{s}}(p) = [a,b]$),
$$
	\mathfrak{d}_{\mathrm{s}} =
	\{(p, E_{\mathrm{s}}(p)): p \in D_{\mathrm{s}}\}
$$
is the set of attribute-set pairs,
with the
mapping $E_{\mathrm{s}}: D_{\mathrm{s}} \to \mathscr{D}$
from the given set $D_{\mathrm{s}}$ of searched discrete attributes
to their desired sets or enums,
($E_{\mathrm{s}}$ associates discrete nonnumerical attribute $p$ a set 
$E_{\mathrm{s}}(p)$), and
$$
	\mathfrak{i}_{\mathrm{s}} 
	= \{(p,l_{\mathrm{s}}(p)): p \in I_\mathrm{s}\}
$$
is the set of searched fields of interest with their desired levels,
with the given mapping $l_{\mathrm{s}}: I_\mathrm{s} \to [-1,1])$.
Note that each of the pairs $(p,R_{\mathrm{s}})$, $(p,E_{\mathrm{s}})$,
$(p,l_{\mathrm{s}})$ can be easily implemented as a table or a hash map.
Analogously, an advertising profile is given by
\begin{equation}
	\mathfrak{a}
	= \mathfrak{n}_{\mathrm{a}} \cup \mathfrak{d}_{\mathrm{a}} \cup \mathfrak{i}_{\mathrm{a}}, 
	\label{eq-advertising-profile}
\end{equation}
where the three sets are defined the same way as in the search case,
with the index `s' (for ``search'') replaced by `a' (for ``advertising'').

\begin{beispiel}
	\label{bsp-intelligent-grid}
	In grid computing\index{grid computing},
	a main matching problem is resource discovery and
	resource allocation \cite{Foster-Kesselman-2004,Prodan-Fahringer-2006}.
	Assume a toy grid	consisting of two resource providers Haegar and Bond,
	and two resource requests by some computational process. 
	In our terminology, Bond and Haegar
	each offer an advertising profile, whereas the requests are represented
	by search profiles. 
	Moreover, in the widely used matchmaking framework \emph{Condor-G}\index{Condor-G}
	\cite{Raman-et-al-1998,Gonzalez-Castano-et-al-2003,
	Lodygensky-et-al-2003,Thain-Livny-2004,Thain-et-al-2005},
	the profiles are called \emph{ClassAds}\index{ClassAd}
	(classified advertisements).
	Let us assume the profiles according to the following tables.
	\begin{center}
	\begin{footnotesize}
	\begin{tabular}[t]{|*{2}{l|}}
		\hline
		\multicolumn{2}{|c|}{\textbf{Search Profiles}}
		\\ \hline \hline
		 owner = 194.94.2.21 &  owner = 194.1.1.3
		\\ \hline \hline
		 CPU $\geq$ 1.6 GHz & memory $\geq$ 2 GB
		\\
		 memory $\geq$ 1 GB &
		\\ \hline
	\end{tabular}
	\end{footnotesize}	\end{center} \begin{center}\begin{footnotesize}
	\begin{tabular}[t]{|*{3}{l|}}
		\hline
		\multicolumn{2}{|c|}{\textbf{Advertising Profiles}}
		\\ \hline \hline
		 owner = bond.cs.ucf.edu & owner = 194.94.2.20 
		\\ \hline \hline
		 CPU $\leq$ 3.6 GHz   & CPU = 2.5 GHz
		\\
		 memory $\leq$ 4.0 GB & memory = 1.0 GB
		\\ \hline
	\end{tabular}
	\end{footnotesize}
	\end{center}
	In each column of a profile there is listed its owner and some 
	attributes and their values.
\end{beispiel}

\begin{beispiel}
	\label{bsp-interest-groups}
	Assume a small social network for pooling interest groups,
	consisting of three persons,
	Alice, Bob, and Carl, who provide search and advertising profiles
	according to the following tables.
	\begin{center}
	\begin{footnotesize}
	\begin{tabular}[t]{|*{2}{l|}}
		\hline
		\multicolumn{2}{|c|}{\textbf{Search Profiles}}
		\\ \hline \hline
		 owner = Alice & owner = Carl
		\\ \hline \hline
		 age $\in$ [20,40] & age $\in$ [20,30]
		\\
		                   & height $>$ 180
		\\ \hline
		 tennis = 1.0 & basketball = 1.0
		\\
		 chess = 0.5  &
		 \\ \hline
	\end{tabular}
	\end{footnotesize}	\end{center} \begin{center}\begin{footnotesize}
	\begin{tabular}[t]{|*{3}{l|}}
		\hline
		\multicolumn{3}{|c|}{\textbf{Advertising Profiles}}
		\\ \hline \hline
		 owner = Alice & owner = Bob & owner = Carl
		\\ \hline \hline
		 age = 25     & age = 26        & age = 31
		\\
		 height = 165 & height = 182    & height = 195
		\\ \hline
		 tennis = 1.0 & tennis = 0.5    & basketball = 1.0
		\\
		 chess = 0.5  & basketball = $-1.0$ &
		\\
		 basketball = 0.5  & &
		 \\ \hline
	\end{tabular}
	\end{footnotesize}
	\end{center}
	In each column of a profile there is listed its owner, some attributes and
	their values, and the fields of interests with their levels.
	For instance, Alice looks for someone between 20 and 40 years of age 
	being enthusiastic in tennis and having some penchant to chess,
	whereas Carl seeks a tall person in the 20's with highest preference for
	basketball.
	Looking at the advertising profiles in this social network, one sees that
	Alice may contact Bob, but Carl cannot find an ideal partner in this community.
	On the other hand, Alice would be a ``better'' partner for Carl than Bob,
	since she is partly interested in basketball.
	Formally Alice's search profile, for instance, is given as follows.
	The sets for the searched attributes and fields of interest are
	\begin{equation}
		N_{\mathrm{s}} = \{\mathrm{age}\},
		\quad
		D_{\mathrm{s}} = \emptyset,
		\quad
		I_{\mathrm{s}} = \{\mathrm{tennis, \ chess}\}
		,		
	\end{equation}
	the mapping
	$R_{\mathrm{s}}$
	is given by
	\begin{equation}
		R_{\mathrm{s}}(p)
		= \left\{ \begin{array}{ll}
			[20, 40] & \mbox{if $p$ $=$ ``age,''}
			\\
			\emptyset & \mbox{otherwise.}
		\end{array} \right.
	\end{equation}
	and the mapping 
	$l_{\mathrm{s}}$ is given by the table
	\begin{equation}
		\begin{array}{c|*{2}{c}}	
			p & \mbox{tennis} & \mbox{chess}
			\\ \hline
			l_{\mathrm{s}}(p)
			& 1.0 & 0.5
		\end{array}
	\end{equation}
	The mapping $E_s$ does not exist since $D_s = \emptyset$.
	To sum up, Alice's search profile is given by
	\begin{eqnarray}
		\mathfrak{s}_{\mathrm{Alice}}
		& \hspace*{-.5ex} = \hspace*{-.5ex} &
		\{(\mbox{age}, [20, 40])\}
		\nonumber \\
		& \hspace*{-.5ex} \cup \hspace*{-.5ex} &		
		\{(\mbox{tennis}, 1.), (\mbox{chess}, .5), (\mbox{basketball}, .5)\}
		, \qquad
	\end{eqnarray}
	Note in particular that $\mathfrak{n}_{\mathfrak{s}, \mathrm{Alice}} = \emptyset$.
	On the other hand, the advertising profiles read
	\begin{equation}
		\mathfrak{A}(\mathfrak{s}_{\mathrm{Alice}})
		= \{\mathfrak{a}_{\mathrm{Bob}},
		  \mathfrak{a}_{\mathrm{Carl}}\}
	\end{equation}
	where
	\begin{eqnarray}
		\mathfrak{a}_{\mathrm{Bob}}
		& \hspace*{-0.5ex} = \hspace*{-0.5ex} &
		\{(\mbox{age}, [26,26]), (\mbox{height}, [182,182])\}
		\nonumber \\
		& \hspace*{-0.5ex} \cup \hspace*{-0.5ex} &
		\{(\mbox{tennis}, .5), (\mbox{basketball}, -1.)\}
		.
	\end{eqnarray}
	\begin{eqnarray}
		\mathfrak{a}_{\mathrm{Carl}}
		& \hspace*{-0.5ex} = \hspace*{-0.5ex} &
		\{(\mbox{age}, [31,31], (\mbox{height}, [195,195])\} 
		\nonumber \\
		& \hspace*{-0.5ex} \cup \hspace*{-0.5ex} &
		\{(\mbox{basketball}, 1.)\}
		.
	\end{eqnarray}
	With the definitions
	\begin{equation}
		s_B =
		(\mathfrak{s}_{\mathrm{Alice}}, \mathfrak{a}_{\mathrm{Bob}})
		, \qquad
		s_C =
		(\mathfrak{s}_{\mathrm{Alice}}, \mathfrak{a}_{\mathrm{Carl}})
		,
		\label{bsp-Alice-solutions}
	\end{equation}
	the search space 
	$
		S 
		= \{\mathfrak{s}_{\mathrm{Alice}}\} 
		\times \{\mathfrak{a}_{\mathrm{Bob}}, \mathfrak{a}_{\mathrm{Carl}}\}
		= \{s_B, s_C\}
	$
	consists of two feasible solutions.
\end{beispiel}

\subsection{A matching degree function}
The \emph{matching degree}\index{matching degree} of a search profile and
an advertising profile is a real number $f$, typically
0 $\leqq$ $f$ $\leqq$ 1, with $f=0$ meaning ``total mismatch'' and $f=1$ meaning
``perfect match.''
In general, the matching degree will be the weighted sum of several
partial matching degrees, one for each property separately.
Moreover, the matching degree of an attribute is calculated in a different
way than the matching degree of a field of interest.
Proposals for these different kinds of matchings are introduced in the
following paragraphs.

\subsubsection{Matching degree of an attribute range}
A function measuring the matching degree of ranges of an attribute has to quantify
how a given advertised attribute stencil $[a_{\mathrm{a}}, b_{\mathrm{a}}]$ fits into the
stencil pattern given by the corresponding range in the search profile.
In case of a numerical attribute, the stencil is given by a closed
interval $[a,b]$ $\in$ $\mathscr{D}$
in case of a discrete-value attribute it is a set or enum $E$.

\bigskip

\begin{figure}[htp]
\centering
	\begin{footnotesize}
	\unitlength1ex
	\begin{picture}(30,16)
		\put(0,0){\includegraphics[width=30ex]{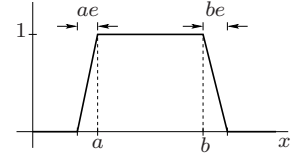}}
		\put(29.5,1.2){\makebox(0,0)[t]{$x$}}
		\put( 1.2,13.0){\makebox(0,0)[r]{1}}
		\put( 9.0,1.0){\makebox(0,0)[t]{$a$}}
		\put(20.8,1.2){\makebox(0,0)[t]{$b$}}
		\put( 7.9,14.6){\makebox(0,0)[b]{$ae$}}
		\put(21.9,14.6){\makebox(0,0)[b]{$be$}}
	\end{picture}
	\end{footnotesize}
	\caption{\label{fig-fuzzy-step}\footnotesize
		The fuzzy step function $h_e(x)$ of Eq.~(\ref{eq-fuzzy-step}).
	}
\end{figure}

\paragraph{Numerical attribute intervals matching.}
To determine the matching degree of a searched value range $[a_{\mathrm{s}},b_{\mathrm{s}}]$ with
a given advwertised value range 
$x \in \mathbb{R}$,
we define the fuzzy step function\index{fuzzy step function}
$h_e(x) = h_e(x;a,b)$ with $a\leqq b$ and
$0<e<1$, as
\begin{equation}
	h_e(x;a,b)
	= \left\{ \begin{array}{ll}
		\frac{x}{ae} - \frac{1-e}{e} & \mbox{if $(1-e)a < x \leqq a$,} \\
		1                            & \mbox{if $a < x \leqq b$,} \\
		\frac{1+e}{e} - \frac{x}{be} & \mbox{if $b < x \leqq (1+e)b$,} \\
		0                            & \mbox{otherwise.}
	\end{array} \right.
	\label{eq-fuzzy-step}
\end{equation}
(Fig.~\ref{fig-fuzzy-step}).
The parameter $e$ is called the \emph{fuzzy level}\index{fuzzy level}.
It denotes the relative length of the fuzzy transition region.
The smaller $e$, the narrower this region, and the more accurate an
advertised attribute value must fit into the searched interval.
In the limit $e \to 0$, the function $h_e$
is the step function, and for $a \to -\infty$ or $b \to \infty$,
it tends to one of the Heaviside step functions $H_b(-x)$ or $H_{\mathrm{a}}(x)$,
respectively.

If, for instance, the searched attribute is ``height $>$ 180''
and an advertised attribute is ``height = 165''
then for a fuzzy level of $e=10\%$, we have
\begin{equation}
	h_{0.1}(165; 180, \infty)
	= \frac{165}{18} - \frac{.9}{.1}
	= 0.1\overline{6},
\end{equation}
i.e., the matching degree is 16.7\%.
Then the matching degree of two numerical ranges 
$[a_{\mathrm{s}},b_{\mathrm{s}}]$ as search range and $[a_{\mathrm{a}},b_{\mathrm{a}}]$ as advertised range 
are given by
\begin{eqnarray}
	\lefteqn{m_n([a_{\mathrm{s}},b_{\mathrm{s}}], \ [a_{\mathrm{a}},b_{\mathrm{a}}]; \ e) =}
	\nonumber \\
	& & \max \left[ h_e(b_{\mathrm{a}}; a_{\mathrm{s}}, b_{\mathrm{s}}), \ h_e(b_{\mathrm{s}}; a_{\mathrm{a}}, b_{\mathrm{a}}) \right]
	.
\end{eqnarray}

\bigskip

\paragraph{Features matching finite sets or enums.}
If the values of specific attribute are constrained to be of a
finite set, or an enum, say $E$ then the matching degree
is determined by the Boolean characteristic function
$\chi_E$ defined by
\begin{equation}
	\chi_E(x) =
	\left\{ \begin{array}{ll}
		1 & \mbox{if $x \in E$,}\\
		0 & \mbox{otherwise.}
	\end{array} \right.
\end{equation}
If the searched attribute, for instance, is
``name $\in$ \{`Smith', `Taylor'\}''
and the advertised attribute is ``name = `Tailor'\/''
then $E$ $=$ \{`Smith', `Taylor'\} and
$\chi_E(\mbox{`Tailor'}) = 0$,
i.e., the matching degree is zero.
Since the owner of an advertising profile can advertise
at most one value for an attribute,
we have
\begin{equation}
	m_d(E,\{x\}) = \chi_E(x).
\end{equation}

\subsubsection{Matching degree of a field of interest}
First we notice that the matching degree as a function of the
levels of interest $l_{\mathrm{s}}$ for the search profile and
$l_{\mathrm{a}}$ for the advertising profile must be asymmetric.
For instance, if $l_{\mathrm{s}}=0$ and $l_{\mathrm{a}}=1$, i.e., the search is
indifferent with respect to the field of interest,
and the advertising profile has $l_{\mathrm{a}}=1$, then the matching degree
should be greater than 0, but if the search requires $l_{\mathrm{s}} = 1$
and $l_{\mathrm{a}}=0$ then the matching degree should be zero.
In the first case, the searcher is indifferent about the field
of interest, in the second case he demands high interest.

\begin{definition}
	\label{def-interest-matching-degree-function}
	An \emph{interest matching degree function}\index{matching degree function}
	is a function $m: [-1,1]^2 \to [0,1]$ such that the following conditions
	are satisfied.
	\begin{equation}
		\begin{array}{|c|ccc|}
			\hline
			(l_{\mathrm{s}},l_{\mathrm{a}})  & (x,x) & (0,\pm 1) & (\pm 1, 0)
			\\ \hline
			m(l_{\mathrm{s}},l_{\mathrm{a}}) &   1   &  \frac12  &     0
			\\ \hline
		\end{array}
		\quad (\forall x \in [-1,1])
		\label{eq-cond-interset-matching-degree}
	\end{equation}
	The first condition 
	expresses the \emph{perfect matching of the diagonal},
	the second the \emph{search indifference},
	and the last the \emph{search necessity}.
%
%
\end{definition}

\noindent
A possible matching degree function is given by
\begin{equation}
	m_i(l_{\mathrm{s}},l_{\mathrm{a}})
	= \max [\varphi(l_{\mathrm{s}}, l_{\mathrm{a}}), \ 0]
	\label{eq-matching-degree}
\end{equation}
where
\begin{equation}
	\varphi(x,y) = 1 - \frac{(c^2-1) (x - y)^2}{c^2 - 2 + (x - c y)^2}
\end{equation}
with
\begin{equation}
	c = \frac{1 + \sqrt7}{2} \approx 1.823
	.
\end{equation}
By construction, $m(l_{\mathrm{s}}, l_{\mathrm{a}})$ satisfies the
conditions in (\ref{eq-cond-interset-matching-degree}) and therefore is
an interest matching degree function.
\begin{figure}[htp]
\centering
	\begin{footnotesize}
	\unitlength1ex
	\begin{picture}(40,25)
		\put(0,0){\includegraphics[width=40ex]{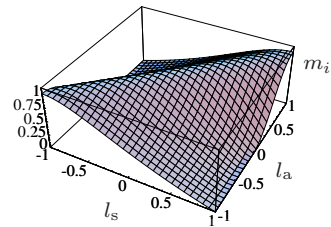}}
		\put(15.0,2.5){\makebox(0,0)[t]{$l_{\mathrm{s}}$}}
		\put(33.0,6.0){\makebox(0,0)[l]{$l_{\mathrm{a}}$}}
		\put(36.0,17.5){\makebox(0,0)[l]{$m_i$}}
	\end{picture}
	\end{footnotesize}
	\caption{\label{fig-matching}\footnotesize
		The matching degree function 
		$m = m(l_{\mathrm{s}},l_{\mathrm{a}})$
		in Eq.~(\ref{eq-matching-degree}).
	}
\end{figure}
It is asymmetric with respect to its arguments, since we have
$m(l_{\mathrm{s}},l_{\mathrm{a}}) \ne m(l_{\mathrm{a}},l_{\mathrm{s}})$ if and only if $l_{\mathrm{s}}^2 \ne l_{\mathrm{a}}^2$.
On the other hand, it is an even function,
i.e., $m(l_{\mathrm{s}},l_{\mathrm{a}}) = m(-l_{\mathrm{s}},-l_{\mathrm{a}})$.

\subsubsection{The total matching degree function}
Putting together all partial matching degrees considered above, we have to
construct a function $f:S \to [0,1]$ as a weighted sum of them.
We notice that any $s \in S$ represents a feasible solution of
the matchmaking problem and has the form
$
	s = (\mathfrak{s}, \mathfrak{a})
$
where $\mathfrak{s}$ is the given search profile (\ref{eq-search-profile})
and
$\mathfrak{a}$ is one of the given advertising profiles (\ref{eq-advertising-profile})
in the network.
Then $f$ defined for each $s\in S$ by
\begin{eqnarray}
	f(s)
	& \hspace*{-0.5ex} = \hspace*{-0.5ex} &
	\sum_{p \in N_{\mathrm{s}}}
	\frac{m_n(R_{\mathrm{s}}(p),R_{\mathrm{a}}(p))}
	{|N_{\mathrm{s}}| + |D_{\mathrm{s}}| + |I_{\mathrm{s}}|}
	\nonumber \\
	& \hspace*{-0.5ex} + \hspace*{-0.5ex} &
	\sum_{p\in D_{\mathrm{s}}} 
	\frac{m_d(E(p), T_{\mathrm{a}}(p))}
	{|N_{\mathrm{s}}| + |D_{\mathrm{s}}| + |I_{\mathrm{s}}|}
	\nonumber \\
	& \hspace*{-0.5ex} + \hspace*{-0.5ex} &	
	\sum_{	p \in I_{\mathrm{s}}}
	\frac{m_i(l_{\mathrm{s}}(p), l_{\mathrm{a}}(p))}
	{|N_{\mathrm{s}}| + |D_{\mathrm{s}}| + |I_{\mathrm{s}}|}
	\qquad
	\label{eq-matchmaking-objective-function}
\end{eqnarray}
where $R_{\mathrm{a}}(p)$ and $E(p)$ denote the attribute ranges of the attribute $p$, 
$l_{\mathrm{a}}(p)$ is the advertised interest level of the 
field of interest $p$,
and the vertical bars $|\cdot|$ embracing a set denote the number of 
its elements.

Thus for the computation of the matching degree,
the attributes and fields of interest 
of the search search profile $\mathfrak{s}$ are leading,
i.e., it is $\mathfrak{s}$ which determines what is
tried to be matched.
The, if an attribute $p$ of the search profile does
not occur in the advertising profile, then the matching degree functions
$m_n(p)$ and $m_d(p)$ vanish by definition.
If, however, a searched field of interest $p \in \mathfrak{i}_{\mathrm{s}}$ does not
occur in the advertised profile, then it is the level $l_{\mathrm{a}}(p)$ which vanishes
by definition.
Note the crucial difference between null values of attributes and null values
of fields of interest in the advertising profile: searched attributes are
mandatory, and at least with respect to this attribute there is a complete mismatch;
if a field of interest, however, does not occur in the advertised profile,
it is indifferent to its owner, but depending on the level of interest in the
search profile, the matching degree may be positive nonetheless.

\bigskip
\noindent
\textbf{Example \ref{bsp-interest-groups} (revisited).}
	For Alice's search space we have the 
	two solutions (\ref{bsp-Alice-solutions}),
	i.e.,
	\begin{eqnarray}
		f(s_B)
		& \hspace*{-0.5ex} = \hspace*{-0.5ex} &
		\frac{m_n([20,40], [26,26])}{3} 
		+ \frac{m_i(1, .5) + m_i(.5, 0)}{3}
		\nonumber \\
		& \hspace*{-0.5ex} = \hspace*{-0.5ex} &	
		\frac{1}{3} + \frac{.5636 + .6308}{3}
		= 0.7315
		\label{bsp-matchmaking-objective-function-Bob}
	\end{eqnarray}
	and
	\begin{eqnarray}
		f(s_C)
		& \hspace*{-0.5ex} = \hspace*{-0.5ex} &
		\frac{m_n([20,40], [31,31])}{3} 
		+ \frac{m_i(1., 0) + m_i(.5, 0)}{3}
		\nonumber \\
		& \hspace*{-0.5ex} = \hspace*{-0.5ex} &	
		\frac{1}{3} + \frac{0 + .6308}{3}
		= 0.5436
		\label{bsp-matchmaking-objective-function-Carl}
	\end{eqnarray}
	Hence Bob's advertising profile has a matching degree of 73.15\% 
	with Alice's search profile, whereas Carl's matches it only by 54.36\%.
\hfill{$\square$}

\bigskip

Notice that the objective function (\ref{eq-matchmaking-objective-function})
is constructed in such a way that each searched item $p$ of a search profile
has equal weight. If, however, each item should have its own weight $w(p)$,
then the objective function (\ref{eq-matchmaking-objective-function-2})
is easily be modified to
\begin{eqnarray}
	f(s)
	& \hspace*{-0.5ex} = \hspace*{-0.5ex} &
	\sum_{p \in N_{\mathrm{s}}}
	\frac{w_n(p)}{w_{\mathrm{tot}}}\
	{m_n(R_{\mathrm{s}}(p),R_{\mathrm{a}}(p))}
	\nonumber \\
	& \hspace*{-0.5ex} + \hspace*{-0.5ex} &
	\sum_{p\in D_{\mathrm{s}}} 
	\frac{w_d(p)}{w_{\mathrm{tot}}}\
	{m_d(E(p), T_{\mathrm{a}}(p))}
	\nonumber \\
	& \hspace*{-0.5ex} + \hspace*{-0.5ex} &	
	\sum_{	p \in I_{\mathrm{s}}}
	\frac{w_n(p)}{w_{\mathrm{tot}}}\
	{m_i(l_{\mathrm{s}}(p), l_{\mathrm{a}}(p))}
	\qquad
	\label{eq-matchmaking-objective-function-2}
\end{eqnarray}
where $w_{\mathrm{tot}}$ is the total sum
\begin{equation}
	w_{\mathrm{tot}}
	= \sum_{p \in N_{\mathrm{s}}} w_n(p)
	+ \sum_{p \in D_{\mathrm{s}}} w_d(p)
	+ \sum_{p \in I_{\mathrm{s}}} w_i(p)
\end{equation}
of all weights 
$w_n$: $N \to \mathbb{R}^+$,
$w_d$: $N \to \mathbb{R}^+$,
$w_i$: $N \to \mathbb{R}^+$.

\section{Discussion}
In this paper, a mathematical model of the matchmaking of search and advertising profiles 
in an electronic social network is proposed. 
Basing on the data structure described by Figure \ref{fig-uml-matchmaking}
and distinguishing between matchmaking of attribute ranges via stencils
and matchmaking of fields of interests via comparison,
the matchmaking problem is formulated as an optimization problem,
with the search space consisting of a fixed search profile and several
advertising profiles as in Eq.~(\ref{eq-matchmaking-search-space})
and the matching degree as its objective function in 
Eq.~(\ref{eq-matchmaking-objective-function}).
The main difficulty is to define a function measuring adequately the
matching degree of two fields of interest and obeying the necessary conditions
listed in Definition \ref{def-interest-matching-degree-function}.
A proposed solution is the function given in Eq.~(\ref{eq-matching-degree})
and depicted in Figure \ref{fig-matching}.
The implementation of a matchmaking service in an electronic social network
basing on this matching optimization is straightforward.

\begin{acknowledgments}
	I am indebted
	to
	Thomas Kowalski, Valerie Rinke, and Volker Weiß
	for valuable discussions.
\end{acknowledgments}




\begin{footnotesize}
\printindex
\end{footnotesize}

\end{document}